# Generalizing Fowler–Nordheim Tunneling Theory for an Arbitrary Power Law Barrier

*Naira Grigoryan, Agata Roszkiewicz, and Piotr Chudzinski\**

**Herein, the canonical Fowler–Nordheim theory is extended by computing the zero-temperature transmission probability for the more general case of a barrier described by a fractional power law. An exact analytical formula is derived, written in terms of Gauss hypergeometric functions, that fully capture the transmission probability for this generalized problem, including screened interaction with the image potential. First, the quality of approximation against the so far most advanced formulation of Fowler–Nordheim, where the transmission is given in terms of elliptic integrals, is benchmarked. In the following, as the barrier is given by a power law, in detail, the dependence of the transmission probability on the exponent of the power law is analyzed. The formalism is compared with results of numerical calculations and its possible experimental relevance is discussed. Finally, it is discussed how the presented solution can be linked in some specific cases with an exact quantum-mechanical solution of the quantum well problem.**

## 1. Introduction

Field-electron emission is a process by which electrons are emitted from a material because of the application of an external electric field. The first successful model for the emission from bulk metals was proposed by Fowler and Nordheim in 1928.[1] Canonical Fowler–Nordheim theory of cold electrons' emission gives a solution for the Schrödinger equation that describes transmission probability computed for a quantum tunneling process through a triangular barrier. The linear dependence of potential is due to the presence of an unscreened external electric field. In more advanced versions, the interaction of an emitted electron with its image potential has also been accounted for. Murphy and Good defined and formulated a theory[2] for the implementation of Fowler–Nordheim equations which became a standard source used in many following research works and educational material. In particular, they have shown that the transmission is given by the exponentiation of the elliptic integral of the first and second kind. Forbes and Deane[3,4] further enhanced the theory by providing a better approximation for elliptic functions and showed that the tunneling process can be described by an underlying ordinary differential equation.

Experimentally, the field-electron emission from the surface of a material is a well-established process. Field-emission electron sources with low-turn-on electric fields, high-emission current density, and good stability have a considerable role in expansive usage as sources for electron microscopes and in special vacuum nanoelectronic equipment.[5] Development of micro- and nanotechnology[6,7] inspired further development of experiments as well as the theory. In a canonical approach, a perfectly flat, conducting surface has been considered. This is justified only when a macroscopic geometrical setting of an experiment is much larger than any microscopic details of the surface. For the smallest devices, that were micro- and nanoengineered in order to reduce their effective emitting surface, this assumption does not hold any longer.

The theory of field emission needs to address these developments and so it remains an active field of research. There is substantial progress being done using numerical methods, using single-particle formalism and dedicated to specific experimental setups. The standard method is to use single-particle formalism of quantum mechanics such as, for example, transfer matrix approach: one uses a numerical solution of Poisson (electrostatic) differential equations to describe the tunneling barrier and then this is integrated numerically. Obviously, the accuracy of calculations depends on the knowledge we have about the surface with a great advantage that it can be easily merged with microscopic knowledge about resonances on the surface[8] which later inspires a merger with ab initio calculations.[9,10] However, the entire sequence of computations need to be repeated if some details of the surface need to be changed; each microscopic case is considered separately. This forms quite exhaustive bulk of work on this subject. When knowledge about the surface is limited, or the surface changes during the emission process, one actually needs to extract surface characteristic details from the tunneling spectrum, which changes the situation. One looks for some general

N. Grigoryan, A. Roszkiewicz, P. Chudzinski
Department of Theory of Continuous Media and Nanostructures
Institute of Fundamental Technological Research
Polish Academy of Sciences
Adolfa Pawinskiego 5b, 02-106 Warsaw, Poland
E-mail: pchudzin@ippt.pan.pl

P. Chudzinski
School of Mathematics and Physics
Queen's University Belfast
University Road, Belfast, NI BT7 1NN, UK

guiding principles that can only be provided by an exact analytic solution for the barrier with a more general shape. The theory then turns toward a continuous effective medium theory for the emitting surfaces with some well-suited parametrization of the broad and appropriate class of potentials.[11] The last, Wentzel–Kramers–Brillouin (WKB), integral still needs to be done numerically.

Here, however, we propose an entirely different approach by proposing a specific, but broad class of potentials for which analytical formula can be given. Apart from its relevance to benchmark numerical results, it is always beneficial to acquire a bit more of a mathematical insight into the subject. Not only will it allow to test numerical results, such analytic formulas are also relevant when tunneling is considered as a part of a larger quantum mechanical problem, when one can simply insert an accurate analytical formula which massively simplifies their treatment. Finally, and most importantly, it allows to grasp the mechanism of tunneling in a more generic case, for example, writing an appropriate differential equation allows to identify universality class of the problem and also to understand the physics of tunneling for the many-body problem.

The article is organized as follows. In Section 2, we define the model. In Section 3, we derive the generalized formula for tunneling in the case of a barrier described by a power law with an arbitrary exponent. In Section 4, using Kemble's improved Jeffreys–Wentzel–Kramers–Brillouin (JWKB) expression for tunneling, we assess the validity of this expression against previous results for the triangular barrier and present results obtained for arbitrary exponents. We study the dependence on the exponent of the power law. We also show tunneling currents for composite surfaces with locally varying work functions. We also make a comparison of our analytical results with a numerical scattering matrix method. Finally, in Section 5, we discuss possible experimental realizations where our theory could apply and build a connection with exact quantum mechanical solutions of the problem. The article is concluded in Section 6.

## 2. The Model

We start with the 1D Schrödinger equation for the tunneling electron

$$-\frac{d^2\Psi(x)}{dx^2} - k_0^2(E - V(x))\Psi(x) = 0 \qquad (1)$$

where $\Psi(x)$ is a 1D wave function, $V(x)$ is the electron potential energy, $E$ is the electron's total forward energy, and $k_0 = \frac{(2m)^{1/2}}{\hbar} \cong 5.123168\,\mathrm{eV}^{-1/2}\,\mathrm{nm}^{-1}$ is a universal constant (see Supporting Information for its derivation). In our case, $V(x)$ will have a specific form of a sum of two opposite power laws. A similar potential has been proposed in ref. [12].

### 2.1. Effective Potential

We see that the entire information about the physics of the problem is inside the effective potential energy experienced by an electron, $V(x)$. The electron image-potential-reduced barrier is shown in **Figure 1**, where $V(x)$ is the electron potential energy, $h$ is the zero-field (tunneling) barrier height ($\omega_0$ is a work function of a given material), and $x_{\mathrm{in}} = x_1$ and $x_{\mathrm{out}} = x_2$ are the inner and outer ends of the barrier (when possible we shall use in this subsection notation from ref. [13] to facilitate comparisons). When power law exponent $\alpha = 1$, it coincides with the exact triangular barrier at perfectly smooth planar interface, while arbitrary $\alpha$ corresponds to the rough surface.

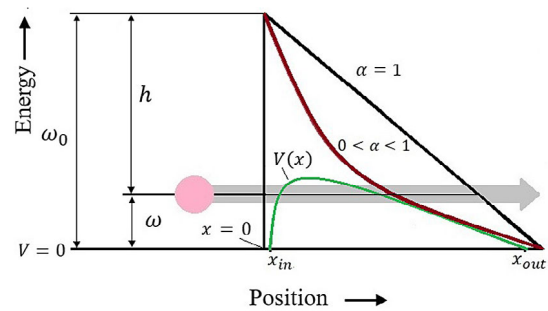

**Figure 1.** Illustration of the image-force-reduced barrier $V(x)$ encountered by a tunneling electron. $\alpha = 1$ corresponds to the exact triangular barrier at a perfectly smooth planar interface, while $0 < \alpha < 1$ corresponds to reduced dimensionality seen by tunneling electrons at a rough surface.

In the past studies, when any attempt for exact analytical expression was made, the $V(x)$ was given by the following expression with $\alpha \equiv 1$

$$V(x) = h - eFx - \frac{e^2}{16\pi\varepsilon_0 x} \qquad (2)$$

Now we generalize this expression for a potential to the following form

$$V(x) = h - eFx^\alpha - \frac{e^2}{16\pi\varepsilon_0 x^\alpha} \qquad (3)$$

where now $x$ is a dimensionless variable which can be constructed either by formal mathematical construct of a surface fractal[12] or by boundary condition set by experiment (We take an electric field $F_{\mathrm{ext}}$ at a small distance from the surface $d_0$ (small enough such that a single power law holds), so $F_{\mathrm{ext}}d_0$ has a unit of energy [eV] and then $x = x_{\mathrm{experim}}[\mathrm{nm}]/d_0[\mathrm{nm}]$). The $h$ is a height of an effective barrier that the tunneling electron is facing. In the absence of any additional exciting force, it reduces to the material characteristic work function $\omega_0$. The theory can be straightforwardly generalized to account for, for example, photon-assisted field emission when electron gains an additional energy $\omega$ before tunneling. Although at present we do not presume to describe any specific realization, this choice of potential may be justified on physical grounds as discussed in detail in Section 5.1.1.

## 3. Solution

In the presence of the image-potential effects, calculations were done in ref. [13], when $\alpha = 1$. They obtained the now widely used elliptic integrals' result given by Equation (13). We are considering a more general case of arbitrary $\alpha$. The evanescent wave function decay is now given by the following, so-called overlap, integral

$$D(F,h) = \exp\left[\frac{(8m_e)^{\frac{1}{2}}}{\hbar}\int_{(h/2eF)(1-s)^{1/\alpha}}^{(h/2eF)(1+s)^{1/\alpha}}\left(h - eFx^\alpha - \frac{e^2}{16\pi\varepsilon_0 x^\alpha}\right)^{\frac{1}{2}}dx\right] \quad (4)$$

The integration limits are $x_{\text{in,out}} = (h/2eF)(1\mp s)^{1/\alpha}$ where $s$ is an auxiliary parameter defined below. By rescaling the integration variable $\xi = (2eF/h)x^\alpha$, we arrive at

$$D(F,h) = \exp\left[-\left(\frac{m_e^{1/2}}{e\hbar}\right)\left(\frac{h^{3/2}}{F}\right)\left(\frac{h}{2eF}\right)^{\frac{1-\alpha}{\alpha}}\left(\frac{1}{\alpha}\right) \times \int_d^c \xi^{\frac{1-\alpha}{\alpha}}\xi^{-\frac{1}{2}}(c-\xi)^{\frac{1}{2}}(\xi-d)^{\frac{1}{2}}d\xi\right] \quad (5)$$

where the last term is an integral sometimes called (up to a prefactor) the Gamow factor or[2,4] principal Schottky–Nordheim barrier function. It reads

$$I(\xi) = \int_d^c \xi^{\frac{1-\alpha}{\alpha}}\xi^{-\frac{1}{2}}(c-\xi)^{\frac{1}{2}}(\xi-d)^{\frac{1}{2}}d\xi \quad (6)$$

We have found the analytic solution of the integral $I(E,F)$, which turns out to be given by a Gauss hypergeometric function

$$I(F,h) = \frac{\pi\alpha d^{\frac{1}{\alpha}-\frac{3}{2}}(2c((\alpha-1)c+d)_2F_1(\frac{1}{2},\frac{3}{2}-\frac{1}{\alpha};1;1-\frac{c}{d}) - \alpha d(c+d)_2F_1(-\frac{1}{2},\frac{3}{2}-\frac{1}{\alpha};1;1-\frac{c}{d}))}{\alpha^2-4} \quad (7)$$

where $I(F,E\equiv h)$ is an *implicit* function of $F$ and $h$ since $c = (1+s)^{1/\alpha}$, $d = (1-s)^{1/\alpha}$ and $s = \sqrt{1-\frac{F}{(\omega_0-\omega)^2}}$.

It is not an accident that the Gamow factor found by us can be expressed by the function that belongs to the $_2F_1$ family, as it has been previously identified[4] that for the case $\alpha = 1$ the defining equation for Fowler–Nordheim tunneling is indeed the Gauss hypergeometric ordinary differential equation (ODE). This can be shown manifestly if we rewrite Equation (7) as follows:

$$I(F,h) = \frac{\pi\alpha d^{\frac{1}{\alpha}+\frac{1}{2}}}{\alpha^2-4}\left(2\left((\alpha-1)\frac{c}{d}+1\right)\frac{c}{d}{}_2F_1\left(\frac{1}{2},\frac{3}{2}-\frac{1}{\alpha};1;1-\frac{c}{d}\right)\right.$$
$$\left. - \alpha\left(\frac{c}{d}+1\right)_2F_1\left(-\frac{1}{2},\frac{3}{2}-\frac{1}{\alpha};1;1-\frac{c}{d}\right)\right) \quad (8)$$

It can now be easily observed that by taking a variable $z \equiv \frac{c}{d}$ (Taking $z \equiv \frac{c}{d}$ implicitly assumes that $c/d$ can vary while $d$ can stay constant which is *mathematically* sound but *physically* hard to reconcile with the fact that the external electric field $F$ is the quantity that one can easily change, which affects equally both $c,d$. However, $c = x_{\text{out}}$ is mostly determined by external fields and one can imagine protocols where only $x_{\text{out}}$ varies keeping $x_{\text{in}}$ intact), the two terms in Equation (8) are in fact two solutions of the hypergeometric ODE

$$(1-z)zw''(z) + w'(z)(\bar{c} - z(\bar{a}+\bar{b}+1)) - \bar{a}\bar{b}w(z) = 0 \quad (9)$$

Precisely, they are solution of the second type in the Kummer list of 24 solutions,[14] namely those of the form $w_{2,1}(z) = {}_2F_1(\bar{a},\bar{b}; \bar{a}+\bar{b}-\bar{c}+1; 1-z)$ and $w_{2,2}(z) = z_2F_1(\bar{c}-\bar{b}+1, \bar{c}+\bar{a}+1; \bar{a}+\bar{b}-\bar{c}+1; 1-z)$ with the following ODE's parameters $\bar{a} = -1/2, \bar{b} = -1/\alpha + 3/2, \bar{c} = -1/\alpha + 1$. Obviously in our solution $w(z)$ there are distinct terms for example the term $((\alpha-1)\frac{c}{d}+1)$ appearing in front of the second solution. This implies that Equation (9) needs to be modified, for instance, by changing the expression in front of $w''(z)$, as $(1-z)z \to (1-z)z/(z+1)$. However, this does not change the number of singular point and it has been proved[15] that every second-order ordinary differential equation with at most three regular singular points can be transformed into the hypergeometric differential equation. Thus WKB problem in general belongs to this class of ODEs. It should be also noted that previously analyzed case $\alpha = 1$ is indeed special, as then some of the terms $\sim(\alpha-1)$ drop, and the differential equation is simpler, namely, it does not contain the "damping" term $\sim w'(z)$. We thus see that a simpler ODE identified in ref. [4] with solution Equation (13) is a special case and for any further quasiclassical analysis of the tunneling mechanism (for instance with time dependence), one should use ODE identified by us, as shown in Equation (9).

## 4. Results

Moving back to observable quantities, we define an overlap quantity $D(F,h)$

$$D(F,h) = \exp\left[-\left(\frac{m_e^{\frac{1}{2}}}{e\hbar_P}\right)\left(\frac{h^{\frac{3}{2}}}{F}\right)\left(\frac{h}{2eF}\right)^{\frac{1-\alpha}{\alpha}}\left(\frac{1}{\alpha}\right)I(F,h)\right] \quad (10)$$

We can now put together the results of the previous section and express the transmission probability in the following way:

$$T = \frac{1}{1+D^{-1}(F,h)} \quad (11)$$

which we shall use in the following to generate the results. This expression, which is a modification of standard WKB[2,16] that suits better concave barriers with a single turning point, has been derived for the first time in ref. [17]. In ref. [18] it was shown that when the top of the tunneling barrier can be approximated by a parabola, then $T = 1/2$ holds for the energy of an electron right on the top of the barrier and Equation (11) (derived thanks to improved, "hydrodynamic" connection formulae)[17] reproduces this result. The formula, Equation (11), has been recently extensively benchmarked against Hill–Wheeler, Wong, and exact quantum mechanical solutions in ref. [19] showing that Kemble's result can be used for the tunneling process both below and above the barrier top. Details are explained in Supporting Information.

### 4.1. Comparison with Elliptic Integrals

When $\alpha = 1$, then analytic WKB solution has been previously obtained[13] as a formula in terms of elliptic integrals

$$D_{\text{past}}(F,h) = \exp\left[-\left(\frac{m_e^{\frac{1}{2}}}{e\hbar_P}\right)\left(\frac{h^{\frac{3}{2}}}{F}\right)I_3(F,h)\right] \quad (12)$$

with

$$I_3 = -\frac{2}{3}(1+a)^{\frac{1}{2}}[E(m) - (1-a)K(m)] \quad (13)$$

$K(m)$ and $E(m)$ are the elliptic integrals of the first and second kind, and $m = \frac{2a}{1+a}$ is the elliptic parameter.

We are now comparing Equation (10) for the general case obtained by us and Equation (12) for the known special elliptic integrals: by definition, they must coincide with each other when $\alpha = 1$. The comparison is made to assess the quality of our new formula since the exact solution is known and widely used in terms of these elliptic integrals. The result of this comparison is presented in **Figure 2**. Indeed, the new function works very well as the surfaces perfectly overlap for all regimes of the external field $F$ and applied photon energy.

### 4.2. Dependence on $\alpha$

Thanks to the fact that we have the generalised formula, now we can explore how the tunneling probability depends on $\alpha$. In **Figure 3** the transmission probability was obtained by our general formula, when $\alpha$ is below 1 and when it is above 1 ($\alpha = 0.3, 0.5, 0.8, 1.2, 1.5, 1.8$). We see that the overall shape of the tunneling probability changes substantially. For the small values of $\alpha$, the transmission probability changes sharply, for example, when $\alpha = 0.3$, there is a very big region where transmission probability is equal to 0, and a very big region where it is equal to 1. For the larger values of $\alpha$ ($> 0.75$), the transmission probability seems to be a varying slope as a function of photon energy capable of shifting the chemical position and thus $h$. For example, when $\alpha = 0.8$, there is a region where transmission probability is equal to 0, then, at a small region of work function or height of the potential, it smoothly elevates from 0 to 1 and there is also a very big region where it is equal to 1. Furthermore, when $\alpha > 1$, an extra edge is appearing.

To investigate this effect further in **Figure 4**a we show the 3D image of the transmission probability dependence on $\alpha$ and $F$, when effective barrier ($\omega_0 - \omega$) is constant and equal to 0.9 eV, while Figure 4b shows the 3D image of transmission probability dependence on $\alpha$ and $F$, when the effective barrier ($\omega_0 - \omega$) is constant and equal to 0.4 eV. In both cases, $\alpha$ changes from 0.1 to 2.5 and the profiles shown in the inset are for the case of $F = 0.5 \text{ V Å}^{-1}$.

There are clearly two regimes: there is a regime where the transmission probability dependence decreases and a regime where it increases as a function of alpha, and there is a critical line between these two regimes, where the quantity is independent of the values of $\alpha$. In Figure 4a,b, two different regimes are shown, we took the cut in the same values of $F$, and these show that in some cases the transmission probability increases as a function of $\alpha$, while in some cases, it decreases.

### 4.3. Contributions from Various Areas of the Surface

We are also able to simulate a surface with spatially varying emission properties. The most general form of emission, for the two-component surface, reads

$$T_{\text{tot}} = s_1 T(F, E; \alpha_1)|_{E=\omega_{01}-\omega} + s_2 T(F, E; \alpha_2)|_{E=\omega_{02}-\omega} \quad (14)$$

where $T(F, E; \alpha_i)$ is given by Equation (11) parameterized by a given $\alpha_i$. For instance, a corrugated surface (see Section 5.1.1) can produce the effect of varying $\alpha$. It is straightforward to realize that the wrapped surface will contain both concave and convex regions. One then expects $\alpha_1 \approx \pi - \gamma$ and $\alpha_2 \approx \pi + \gamma$. In order to match experimental situation, one can assume different area proportions of these regions (and possibly their different densities of states) and see how the total emission current changes. Here, for illustrative purposes, we shall take the simplest situation, where both contribute equally, that is, $s_1 = s_2 = 1/2$.

In **Figure 5**, the transmission probability from the composite surface is presented. We consider two distinct situations: either the work functions $\omega_{0i}$ of the two components are equal or different from each other. In Figure 5a, concave and convex regions have $\alpha_1 = 0.8$ and $\alpha_2 = 1.8$ respectively and $\omega_{01} = \omega_{02} = 1$ eV. Here $T(E)$ is a monotonically increasing function of $F$, as it was expected from the continuously increasing Gamow factor. There are only small, subtle differences in comparison with an average $\bar{\alpha} = 1.4$ case: the intermediate $T(E)$ range is smaller, its increase is smoother and more concave. As regards Figure 5b, where $\alpha_1 = 0.8$, $\omega_{01} = 1$ eV and $\alpha_2 = 1.8$, $\omega_{02} = 0.9$ eV, we do have a double-edge structure which comes from the fact that we have two different work functions of the surface.

A situation when local work functions are different on different areas on the surface may be induced by different crystallographic orientations on the surface (hence differences in surface electric dipole moment), but can also be induced by many-body effects (e.g., charge density waves (CDW) formation). We see that the double edge is particularly visible in the regime of small $\omega$ (deep inside the well) and large enough external field. In our calculations we have also found that the double edge is well pronounced when $\alpha_1 < \alpha_2$ and $\omega_{01} > \omega_{02}$, but much less pronounced in the opposite case when $\alpha_1 < \alpha_2$ and $\omega_{01} < \omega_{02}$.

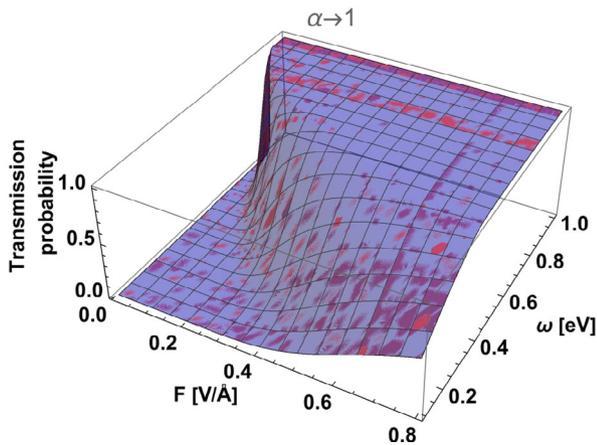

**Figure 2.** Transmission probability as computed by our general formula, Equation (7) (case when $\alpha = 1$) in red and previous result that has been expressed as the elliptic integrals, Equation (13), in blue. Perfect overlap of the two opalescent surfaces results in a single surface with purple color.

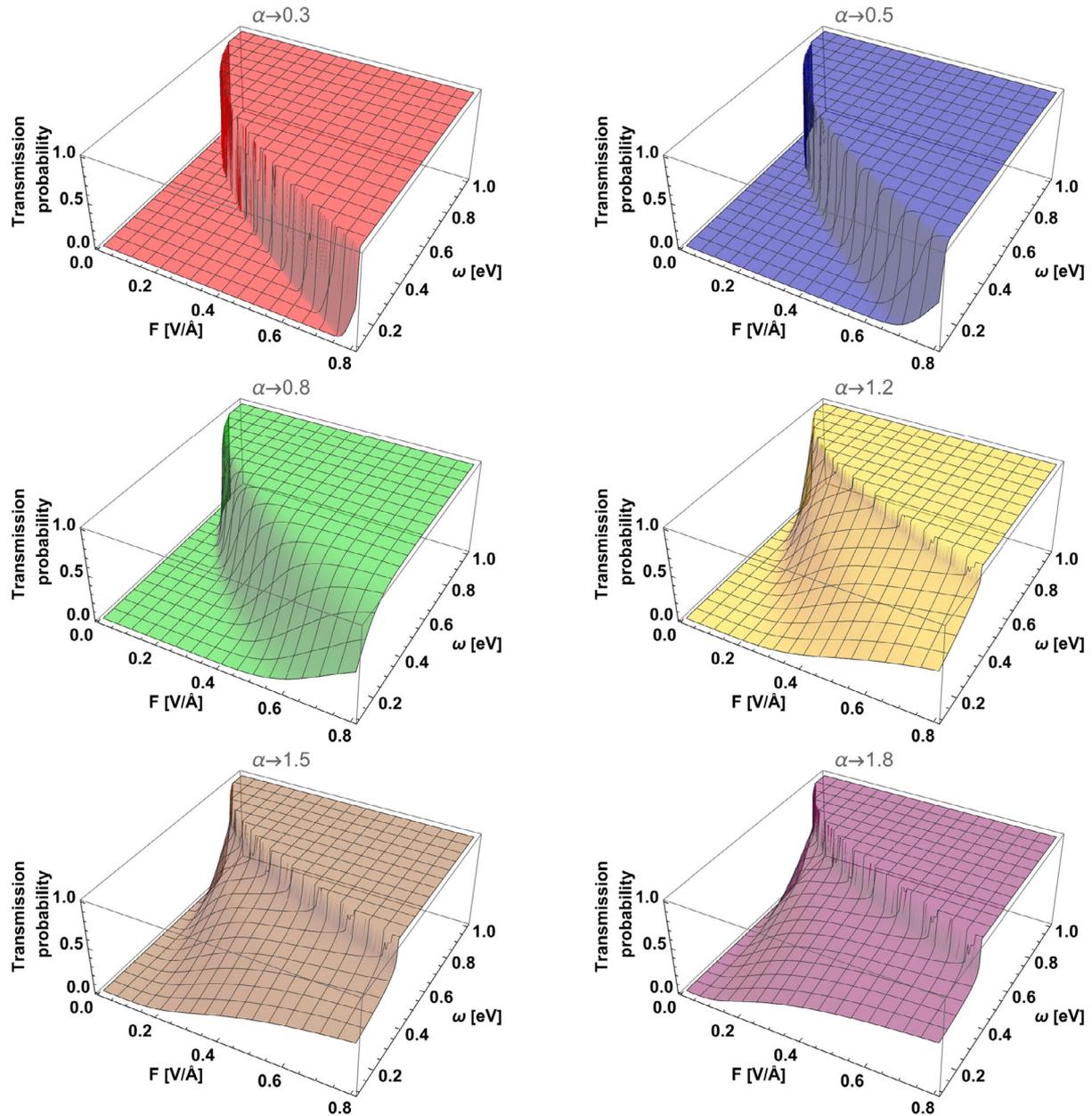

**Figure 3.** Transmission probability calculated by the new general formula expressed by hypergeometric function for the cases (left to right, top to bottom) when $\alpha = 0.3$ (red), $\alpha = 0.5$ (blue), $\alpha = 0.8$ (green), $\alpha = 1.2$ (yellow), $\alpha = 1.5$ (brown), $\alpha = 1.8$ (purple).

Obviously, in the above given expression, we are taking the constant featureless density of states (DOS) for the entire 2D surface. In reality, this needs to be substituted by independently computed material specific data, for example, calculated using ab initio methods. Nevertheless, our formula plays an important role as any DOS has to be multiplied by our result to obtain a measurable quantity. It is thanks to this that we will be able to distinguish what is the proportion of each phase, with different $\omega_{0i}$, on the surface and perhaps even which phase sits on the convex and which on the concave structures.

### 4.4. Comparison with Numerical Methods

The problem of tunneling through a barrier is frequently solved numerically, using transfer matrix or scattering matrix techniques. We shall here make a small comparison between those methods and our analytic formula using a light–matter equivalence.[20] The numerical analysis is based on the 2D Fourier modal method with the implementation of the scattering matrix algorithm and proper factorization rules, extended to multilayer structures.[21] First, we should note that any transfer matrix method improves with increasing number of subsystems and ultimately becomes an ideal approximation in the limit where one divides the potential into an infinite number of slabs, that is, there is an infinite multiplications of matrices involved. Of course, this is impossible from the numerical viewpoint, but it has been shown[22] that this limit is actually *equivalent* to the WKB solution, *provided* the off-diagonal reflection coefficients are neglected.

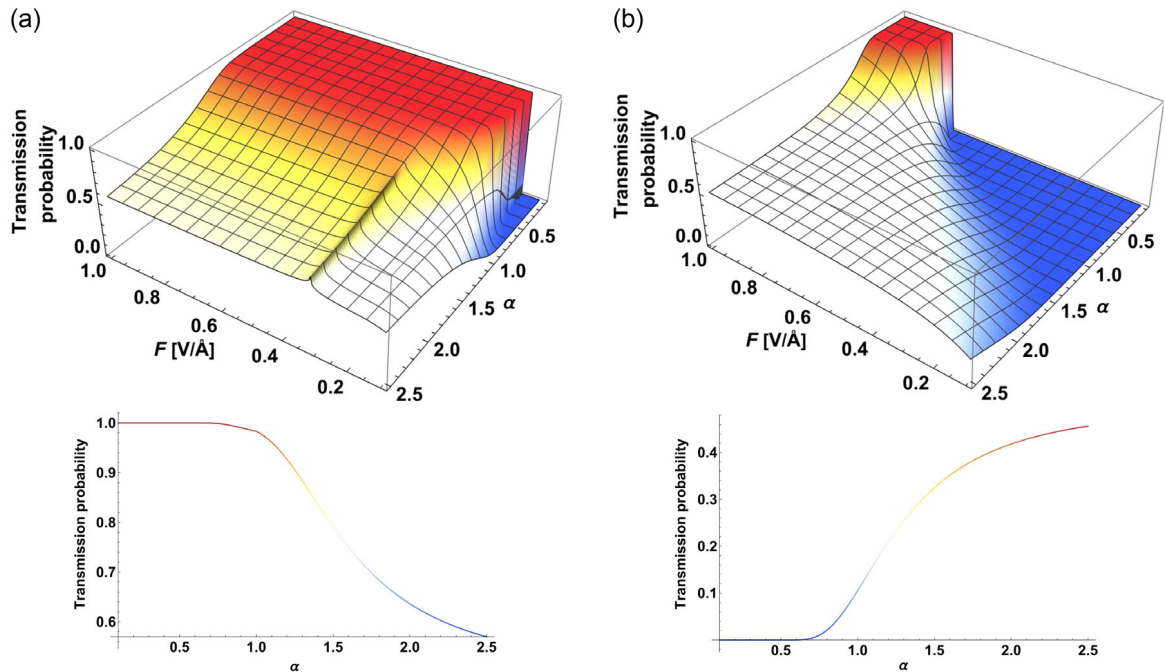

**Figure 4.** a) The transmission probability dependence on $\alpha$, when effective barrier is a) 0.4 eV, b) 0.9 eV, and $\alpha$ changes from 0.1 to 2.5. Profiles in the insets are shown in the case of $F = 0.5\,\mathrm{V\,\mathring{A}^{-1}}$.

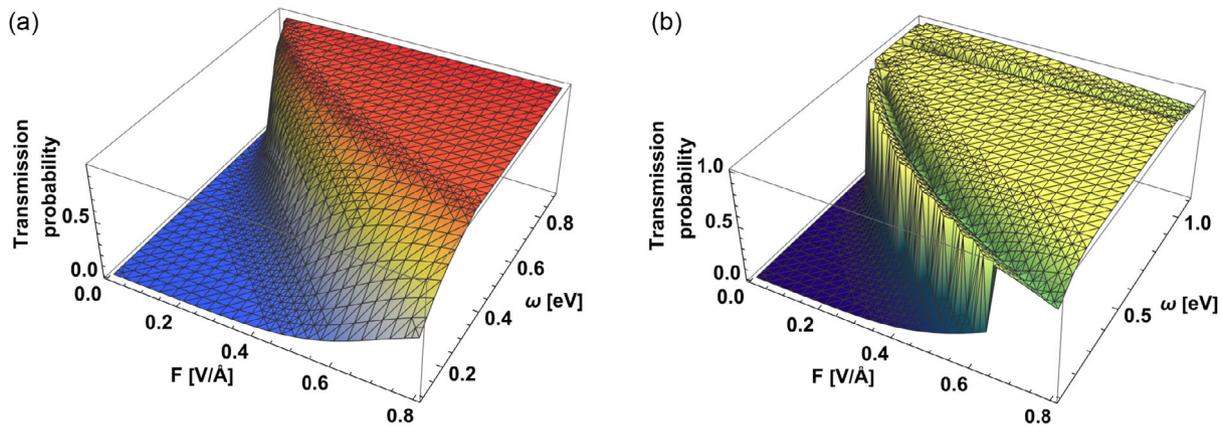

**Figure 5.** The transmission probability with the composite surface computed by Equation (14), when a) $\alpha_1 = 0.8$, $\alpha_2 = 1.8$, and $\omega_{01} = \omega_{02} = 1$ eV and b) $\alpha_1 = 0.8$, $\omega_{01} = 1$ eV and $\alpha_2 = 1.8$, $\omega_{02} = 0.9$ eV.

From this fact, two conclusions can be drawn. First, the smoother the shape of the potential, the smaller the internal reflectivity, and the better WKB will work. This implies that WKB should not be used for tunneling through a potential barrier hosting resonant-level subsystems, for example, quantum dot or adatom on the surface or to be more precise, WKB could be used only for each subsystem separately and then their transfer matrices combined. Second, in order to make a comparison with the scattering matrix approach, one needs to take into account multiple events when the carrier is scattered back (twice) into the barrier and only after that is transmitted. At present all scattering formalism by default includes these corrections, so does the approach introduced in ref. [23], with which we compare. Fortunately, from our result, one can easily deduce the reflection coefficient as well, since $R(F, \omega) = 1 - T(F, \omega)$. Thus we can construct a series of higher-order transmission terms $T = T_{\text{WKB}} + R^2 \Upsilon^2 T_{\text{WKB}} + R^4 \Upsilon^4 T_{\text{WKB}} + \ldots$ which can be resummed as a geometric series. Above we introduced an additional factor $\Upsilon$ which describes the wave function decay as it propagates. For the waves above the barrier, or very close to the barrier top, we expect $\Upsilon = 1$, but deep inside the barrier, for evanescent wave solution we expect $\Upsilon \to 0$ and there indeed pure WKB works well (see **Figure 6**). In the plot below we show a comparison of the numerical method with zeroth-order WKB result and such a resummation with $\Upsilon = 1$. We observe in **Figure 7** that the resummation of the geometrical series allows to establish quite good correspondence between the numerical and analytical methods for larger $\omega$. In the range of the highest transmissions the numerical method reveals the presence of oscillations that are due to the quantum interference effect.

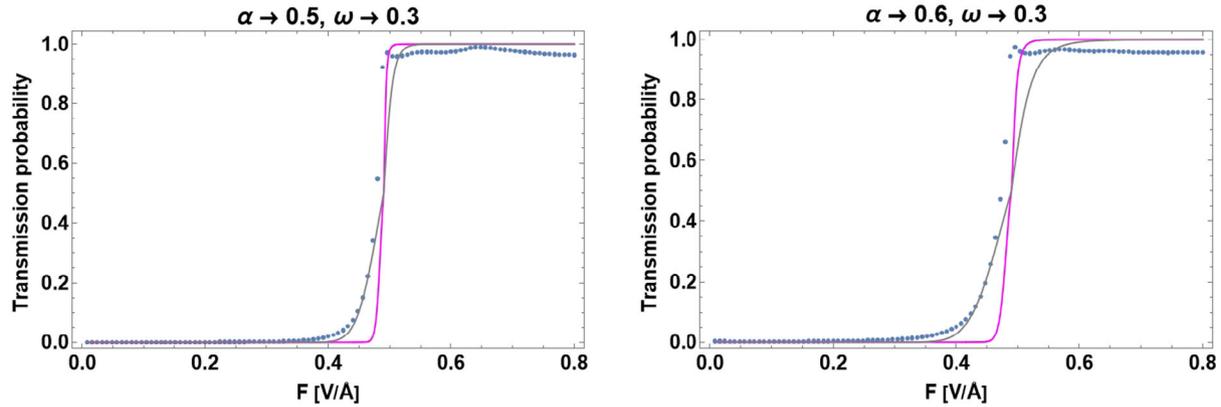

**Figure 6.** The transmission probability comparison between numerical and analytical solutions: the blue points indicate the scattering matrix method calculation with 20 slabs, the pink line indicates the WKB solution considering the internal reflections, and the gray line shows the pure WKB solution.

Our method cannot capture these; however, an extension in this direction could be in principle possible. In any case, the effect turned out to be tiny for our power law barriers. There are also advantages of the exact analytical WKB formula: since the result is given as a single special function (i.e., there is no demand for a final summation of a series or a convolution of special functions), its computational cost is negligible, comparable with a few matrix multiplications. For the numerical method we observe that the result becomes stable when ≈20 slabs are taken (see **Figure 8**); moreover, one needs to be prudent with numerical stability of the matrix multiplication. Finally, it is a challenge to compute numerically data that would generate an analogue of Figure 4, which is particularly relevant for systems where $\alpha$ may change in the course of experiment, such as sputtering phenomena.

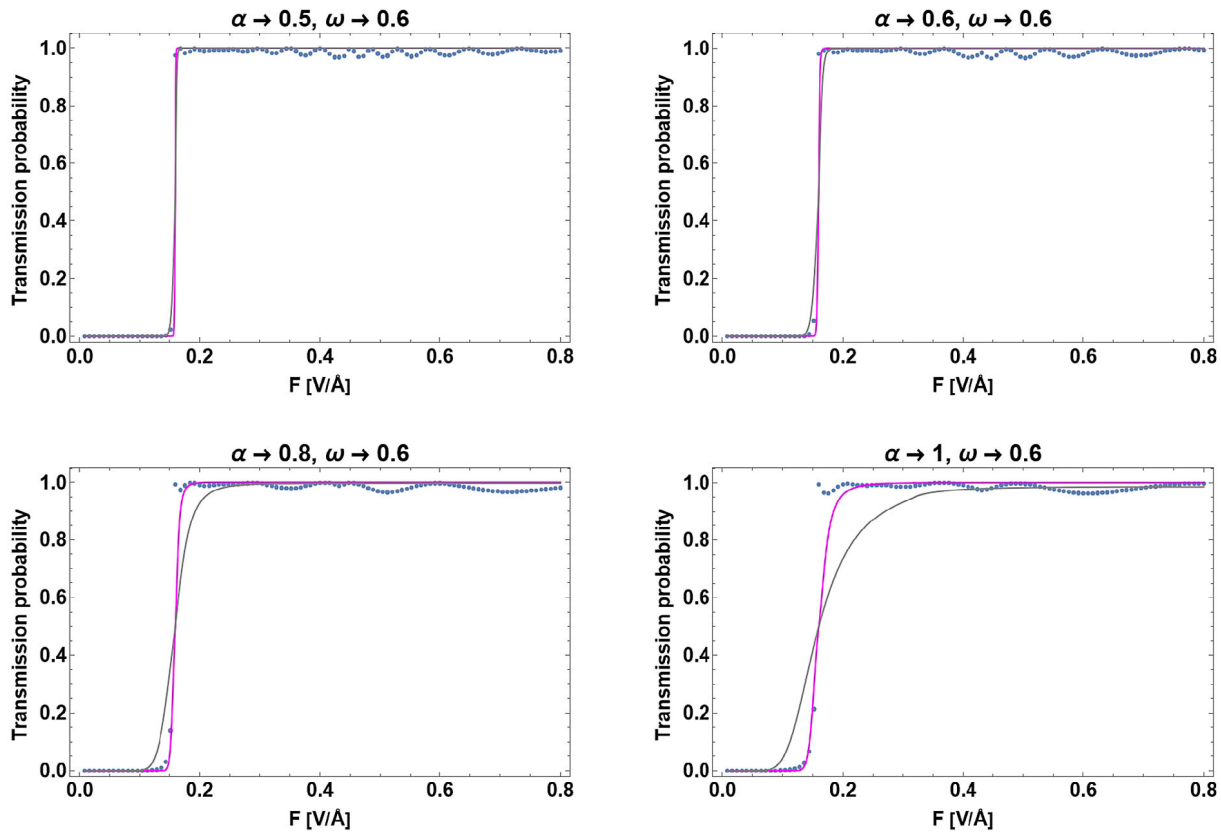

**Figure 7.** The transmission probability comparison between numerical and analytical solutions: the blue points indicate the scattering matrix method calculation with 20 slabs, the pink line indicates the WKB solution considering the internal reflections, and the gray line shows the pure WKB solution.

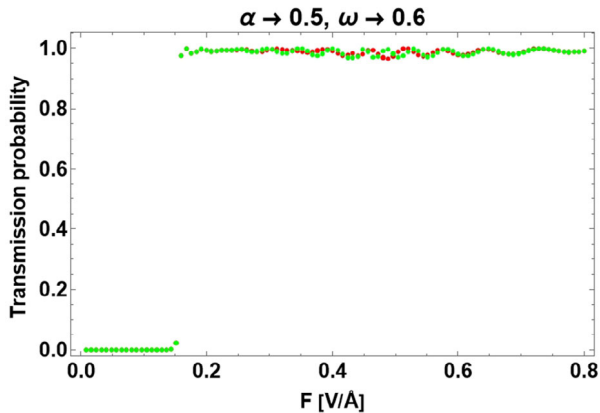

**Figure 8.** The transmission probability calculated by scattering matrix method: red points indicate calculation with 5 slabs and green ones with 20 slabs.

## 5. Discussion

### 5.1. Experimental Relevance

Due to its technological importance, the research field of field-emission applied theory is extremely broad. While the largest amount of work is done purely numerically, we ought to point out works where an analytical approach to external potential has been used. These were actually numerous; a literature search solely for the case of the ellipsoidal tip reveals cases, where the potential has been approximated either as an ellipsoid[24] (with logarithmic dependence $V(r)$) or a power series in terms of Legendre functions.[25] An extensive list of such analytical potentials has been given in ref. [26]. The need for these works was due to the fact that the authors wished to have an insight in how parameters of the potential affect the transmission, hence a similar motivation to ours. However, the last WKB integral was always performed numerically. In our work, we *assume* a given shape of the potential to get a closed analytical formula for the final result, the transmission itself. We cannot presume to achieve similar detail description of potential as the aforementioned works, dedicated to this subject. However, we wish to point out a few arguments showing that our assumption is actually plausible, certainly not unrealistic.

#### 5.1.1. Metallic Surfaces

If the surface is metallic, then we can use knowledge from basic electrostatics to predict power law behavior in its vicinity. For a wrapped metallic surface electrostatic, textbooks [e.g., Jackson,[27] Chap. 2.11] tell us that the power law of external potential depends on the angle of each metallic corner. In general, one expects the following power law of electric field: $x^{\pi/\beta}$. For elevated areas, with corner angles larger than $\pi$ we then expect a behavior like $r^\alpha$ with $\alpha < 1$, hence faster decay of external field (but definitely as a power law). In the extreme case of sharp tip, we have $\beta = 2\pi \Rightarrow \alpha = 1/2$, which is the smallest possible value of $\alpha$ for the entirely metallic 2D surface. On the other hand, for lower (convex) areas, the corner angles are smaller than $\pi$ and so we now expect a behaviour like $r^\alpha$ with $\alpha > 1$ for the external field. The image potential, the other term in Equation (3), will also change depending on surface corrugation: when an electron approaches narrow, fine corners, it is necessary to introduce more (artificial) image charges. This corresponds to dipolar and quadrupolar moments and so the potential diverges faster than $1/r$. Overall the spatial dependence of both terms, the external field and image potential, changes, but in both cases these changes can be potentially captured by our formula Equation (3).

#### 5.1.2. Other, Nonmetallic Surfaces

In the above given situation, we considered a purely metallic, but corrugated surface, where the elementary electrostatic solution is readily accessible. In general, we can assume two coexisting phases on the surface: a dielectric and a metal. The details will depend on the system under consideration, but a few general remarks can be made. For the dielectric, for example, a layer of an oxide gradually covering all available areas, it is likely that we shall encounter a fractal structure, with a partial (Hausdorf) dimension which upon averaging will lead to varying dielectric constant $\varepsilon(x)$ that goes down to $\approx 1$ as we move toward the outside medium. In the effective medium approximation, the local $\varepsilon(x)$ will be proportional to the surface coverage of the dielectric (within a given cross section) which leads to a power law dependence $\varepsilon(x) \approx x^b$. Naturally, this shall affect both ingredients in the interaction part of the Hamiltonian: the image potential will be rescaled by a factor $1/\varepsilon(x)$, while the external potential by a factor $\varepsilon(x)$. This simply means that in the absence of corrugation, but in the presence of the oxide layer, one should take $\alpha = 1 + b$.

The reasoning can be also extended to the case when the kinetic energy part of the Hamiltonian is varying in the direction perpendicular to the surface. As shown in ref. [12], by a proper substitution of the $x$-variable, we are able to rewrite the system with varying content of metallic phase (again as a fractal) or a system with a varying mass $m(x)$ into the form Equation (3); hence, our formalism should be also applicable.

### 5.2. The Case of $\alpha = 2$ and Connection with an Exact Quantum Solution

We observe that the expression we found, Equation (7), suffers from a singularity when $\alpha = 2$. Numerical studies allowed us to check that any small deviation $\pm \varepsilon \to 0$ removes instability and produces a sensible result. Hence only the case $\alpha = 2$ requires a separate insight. First we note that the $1/x^2$ potential well is known[28] to be a pathological case where the well-defined quantum states can be defined only upon imposing an UV cutoff in the problem and upon adding an extra-boundary condition for a derivative.

The need to add the derivative boundary condition suggests that one needs to go beyond WKB to remove the singularity. Quite remarkably, when $\alpha = 2$, Equation (7) actually *simplifies* as now the second index of hypergeometric functions $_2F_1$ equals 1, and so it simplifies to $_1F_1$ otherwise known as Kummer and Triconi confluent hypergeometric functions (since the third index is also equal to 1, it could be simplified further to a binomial, but we shall refrain from that because of reasons that will become clear momentarily). On the other hand, by a substitution of variables $x^2 \to \xi$ we write the following Schrödinger equation

$$\xi \nabla^2 \psi + \nabla \psi + (1/\xi + f\xi - \nu) = 0 \qquad (15)$$

Here the presence of the external force $f$ introduces a natural energy scale in the problem. It is not a pathological case any longer and actually it admits an exact analytic solution. Remarkably, it can be written as a combination of Kummer and Triconi confluent hyperbolic functions (Some languages of symbolic calculations return the solution in terms of generalized Laguerre polynomial, but the latter one is equivalent to the Kummer function) in the following form $\psi(a) = \exp(-\iota k_0 a)(1-a) \, _1F_1\left(\pm 1/2 + \iota \frac{a}{1-a}, 1 + 2\iota \frac{a}{1-a}, \frac{2a}{1-a}\right)$. The form is remarkably similar to our Equation (7) when $\alpha = 2$.

To elaborate on this comparison, one can expand the exponential function (for small argument) in Equation (11) and compare it with an exact quantum mechanical tunneling expression $T \approx 1/[2 + 2\exp(\iota k_0 x_{in} - x_{out})((f(x_{in}/x_{out}) - g(x_{in}/x_{out}))/(f'(x_{in}/x_{out}) - g'(x_{in}/x_{out})))]$ where we used energy current conservation and $f, g$ are the two linearly independent solutions of Schrödinger equation, the Kummer and Triconi functions in our specific case. We see that for $\alpha = 2$ we can make a *full* connection with quantum mechanical solution provided we add an extra phase shift in the first two indexes of $_1F_1$ functions and that we include the $f', g'$ derivatives in the denominator, as they were naturally neglected in WKB. Please note that the derivative of Equation (7) will naturally contain the problematic $1/(\alpha^2 - 4)$ factor and so in the full solution the singularities in the numerator and denominator will cancel each other.

In summary, this detailed study of $\alpha = 2$ case allowed us to make a direct link with the exact quantum mechanical solution of the problem. This not only justifies the Kemble version of JWKB, but also shows that making a generalization of tunneling expression and writing it in terms of hypergeometric $_2F_1$ functions is very useful from the fundamental viewpoint. It remains to be shown whether the corrections to JWKB identified here can be applied for any value of $\alpha$ (the general solution of the quantum mechanical problem for an arbitrary $\alpha$ is not available at present), but it can serve as a relevant benchmark for future experiments and numerical studies.

## 6. Conclusion

In conclusion, the main result of our article is to derive in the JWKB approximation the exact analytic tunneling formula for the barrier described by a fractional power law. The formalism incorporates the external electric field as well as interaction with an image charge left behind in the surface. Our potential is quite specific, as it involves two power laws of precisely opposite exponents, but the existence of the exact closed analytic solution for the transmission $T(E)$, expressed in terms of an easy-to-evaluate single special function, is nevertheless a remarkable result We showed that quite rich tunneling spectrum is possible for the composite surface, paving the way for future analytical modeling of experimental findings. Finally, we showed that our result is general enough to build a connection with some cases of the exact quantum mechanical result.

## Supporting Information

Supporting Information is available from the Wiley Online Library or from the author.

## Acknowledgements

The authors would like to thank Professor Richard Forbes for his valuable comments to the manuscript. P.C. acknowledges financial support of EU through MSCA (grant no. 847639).

## Conflict of Interest

The authors declare no conflict of interest.

## Data Availability Statement

The data that support the findings of this study are available from the corresponding author upon reasonable request.

## Keywords

Gauss hypergeometric functions, cold-electron emission, nanostructures, Wentzel–Kramers–Brillouin approximation

Received: February 3, 2023
Revised: March 1, 2023
Published online: